\documentclass[article,prl,aps,twocolumn,showpacs,floatfix,superscriptaddress]{revtex4-1}
\usepackage{amsmath}
\usepackage{amsthm}
\usepackage{amsfonts}
\usepackage{bbm}
\usepackage{color}
\usepackage{graphicx}
\usepackage{dcolumn}
\usepackage{array}
\usepackage{float}
\usepackage{supertabular}
\usepackage{longtable}
\usepackage{txfonts}
\usepackage{wasysym}
\usepackage{cases}
\usepackage[T1]{fontenc}
\usepackage[latin1]{inputenc}
\usepackage{amssymb}
\usepackage[usenames,dvipsnames]{xcolor}
\usepackage{amstext}
\usepackage{latexsym}
\usepackage[colorlinks=true,citecolor=Blue,linkcolor=RubineRed,urlcolor=Blue]{hyperref}
\usepackage{enumitem}

\usepackage{amsfonts}
\usepackage{epsfig}
\usepackage{dsfont}
\usepackage{mathrsfs}
\usepackage{arydshln,leftidx,mathtools}
\begin{document}
\title{Exact ground states of  quantum many-body systems under confinement}
\author{Adolfo del Campo}
\affiliation{Donostia International Physics Center,  E-20018 San Sebasti\'an, Spain}
\affiliation{IKERBASQUE, Basque Foundation for Science, E-48013 Bilbao, Spain}
\affiliation{Department of Physics, University of Massachusetts, Boston, MA 02125, USA}
\affiliation{Theory Division, Los Alamos National Laboratory, MS-B213, Los Alamos, NM 87545, USA}

\def\L{{\rm \hat{L}}}
\def\q{{\bf q}}
\def\l{\left}
\def\r{\right}
\def\te{\mbox{e}}
\def\d{{\rm d}}
\def\t{{\rm t}}
\def\K{{\rm K}}
\def\N{{\rm N}}
\def\H{{\rm H}}
\def\la{\langle}
\def\ra{\rangle}
\def\om{\omega}
\def\Om{\Omega}
\def\vep{\varepsilon}
\def\wh{\widehat}
\def\tr{{\rm Tr}}
\def\da{\dagger}
\def\iz{\left}
\def\zi{\right}
\newcommand{\beq}{\begin{equation}}
\newcommand{\eeq}{\end{equation}}
\newcommand{\beqa}{\begin{eqnarray}}
\newcommand{\eeqa}{\end{eqnarray}}
\newcommand{\intf}{\int_{-\infty}^\infty}
\newcommand{\into}{\int_0^\infty}

\begin{abstract}
Knowledge of the ground state of  a homogeneous quantum many-body system can be  used to find the exact ground state of a dual  inhomogeneous system with a confining potential. For the complete family of parent Hamiltonians with a  ground state of  Bijl-Jastrow form in free space, the dual system is shown to include a one-body harmonic potential and two-body long-range interactions. The extension to anharmonic potentials and quantum solids with Nosanov-Jastrow wavefunctions is also presented. We  apply this exact mapping to construct eigenstates of  trapped systems from free-space solutions  with a variety of pair correlation functions and interparticle interactions.

\end{abstract}

\maketitle

Exact solutions play an important role in physics. Solvable models often bring novel insights, they can serve as a test-bed for physical theories and a starting point for new approximations, and help benchmarking numerical methods. Solvable models are often integrable. In the classical domain, this requires the existence of  a number of conserved quantities   equal   to (or even greater than) that  of degrees of freedom. In the quantum domain, integrability is often associated with scattering without diffraction, encoded in the Yang-Baxter equations \cite{Sutherland04}. Quantum  integrable systems are generally homogeneous. For quantum many-body systems of continuous variables (e.g., those describing quantum gases and liquids), known exact solutions are typically associated with the absence of an external potential. In the one-dimensional case, an exact treatment is possible via the Bethe ansatz approach, which expresses the wavefunction as a superposition of plane waves \cite{KBI97,Takahashi99,Gaudin14}. This reduces the possible settings to free space (no external one-body potential), translationally-invariant ring geometries with periodic boundary conditions, or settings with hard-walls such as mirrors and box-like traps \cite{Gaudin71,Batchelor06}. 

Known solutions for systems in the presence of an external confining potential that varies gradually in space are rare. Among them, the most prominent case is perhaps that of the rational Calogero-Sutherland gas, describing one-dimensional bosons with inverse-square interactions in a harmonic trap \cite{Calogero71,Sutherland71,Sutherland04}. The latter includes as well hard-core bosons, the so-called Tonks-Girardeau gas, which was first studied in the continuum \cite{Girardeau60} and then in a harmonic trap \cite{GWT01} (as well as ring and box-like traps). While some tools, like the Bose-Fermi duality and anyon-fermion mapping \cite{Girardeau06,delcampo08} or the generalizations to mixtures \cite{Ujino98,GirardeauMinguzzi07}, increases the variety of models, these can be considered all part of the same family.

In the study of nonlinear physics, variants of integrable solutions in free space have been found in inhomogeneous systems. However, as pointed out by Kundu, these are related to homogeneous systems  via gauge, scaling and coordinate transformations \cite{Kundu09}.
An analogue approach has proved  successful in the study of ultracold fermions in the unitary limit \cite{Giorgini08}. In this context, it has been possible to relate universal zero-energy states in free-space, exhibiting scale invariance, with eigenstates of the same system embedded in a harmonic trap \cite{tan2004short,WernerCastin06,Castin12}.

 Knowledge of the ground state of a many-body quantum system is of great importance, as it structure determines the nature of  low-lying excitations built upon it. In addition, as the ground state  includes important correlations among particles, it may be the case that excitations are free, see e.g. \cite{Kawakami93,Gurappa98,Gurappa99,Sutherland04}. In some cases, the ground state determines the structure of  the complete spectrum, as in the Calogero-Sutherland gas \cite{Vacek94,Kawakami94,Sogo94,Sogo96,Gurappa99}. In quasi-solvable models, knowledge of the ground state allows to construct towers of excited states that span part of the spectrum.  An example of this type is the Jain-Khare model involving bosons with two-body and three-body inverse-square interactions among nearest and next-nearest neighbors, respectively \cite{JainKhare99,Auberson01,Basu-Mallick01,Ezung05,Enciso05}. The same holds true for the truncated Calogero-Sutherland gas with interactions restricted to a number of neighbors \cite{Pittman17,Tummuru17}.

In this work, we  introduce a general construction to find quasi-solvable models describing particles confined in a harmonic trap and subject to  long-range pair-wise interactions. We find the complete family of models describing indistinguishable bosons in a harmonic trapped with a ground state of Bijl-Jastrow form. This construction is then generalized to anharmonic external potentials and the description of quantum solids with ground-state wavefunctions of Nosanov-Jastrow form. The power of our approach to identify  (quasi)-solvable and (quasi)-integrable models is unveiled by applying it in a number of scenarios.

{\it Mapping.---}
Consider a many-body problem in the absence of a trapping potential and satisfying
a time-independent Schr\"odinger equation.
\beqa
\hat{H}_0|\Phi_n\ra=E_n|\Phi_n\ra, 
\eeqa
with a generic many-body Hamiltonian in one spatial dimension
\beqa
\hat{H}_0=-\frac{\hbar^2}{2m}\sum_{i=1}^{N}\frac{\partial^2}{\partial x_i^2} +  \sum_{ i < j} v(x_i,x_j)+ \sum_{ i < j<k} w(x_i,x_j,x_k)+\dots\nonumber
\eeqa
in which $v(x,y)$ accounts for the two-body interactions, $w(x,y,z)$ for three body interactions, and so on. In particular, we denote the wavefunction of the ground state of $\hat{H}_0$ by $\Phi_0$  and its eigenenergy by $E_0$.

We are interested in the embedding of this system in a harmonic trap so that the total Hamiltonian reads
\beqa
\label{ParentH}
\hat{H}=\hat{H}_0+\sum_{i=1}^N\frac{1}{2}m\om^2x_i^2+V(x_1,\dots,x_N),
\eeqa
where $V(x_1,\dots,x_N)$ is an additional interaction potential that will be required for consistency.
As an ansatz for the ground state in the presence of the harmonic trap we try the wavefunction
\beqa
\label{PsiPhiTrap}
\Psi_0(x_1,\dots,x_N)=\exp\left(-\frac{m\om}{2\hbar}\sum_{i=1}^Nx_i^2\right)\Phi_0(x_1,\dots,x_N).
\eeqa
To derive the parent Hamiltonian (\ref{ParentH}), we compute the action of the  kinetic energy operator on $\Psi_0(x_1,\dots,x_N)$. It is not difficult to show that
\beqa
\label{nearTISE}
& & \left[\hat{H}_0+\sum_{i=1}^N\frac{1}{2}m\om^2x_i^2\right]\Psi_0-\hbar\om e^{-\frac{m\om}{2\hbar}\sum_{i=1}^Nx_i^2}
\sum_{i=1}^Nx_i\partial_{x_i}\Phi_0\nonumber\\
& & =\left(E_0+\frac{N\hbar\om}{2}\right)\Psi_0.
\eeqa
We are interested in rewriting this equation as a standard many-body Schr\"odinger equation, without velocity-dependent interactions. As we shall see, this  is possible in a number of cases.
For instance, it is known that universal zero-energy states $\Psi_0$ exhibiting scale invariance are eigenstates of the dilatation operator $\hat{D}=\sum_{i=1}^Nx_i\partial_{x_i}$, satisfying $\hat{D}\Psi_0=\nu \Psi_0$ with eigenvalue $\nu$. In such a case,  Eq. (\ref{nearTISE}) does indeed reduce to a many-body Schr\"odinger equation.
This is the strategy used  to find eigenstates of a unitary Fermi gas in a three-dimensional harmonic trap \cite{tan2004short,WernerCastin06,Castin12}.

In what follows, we shall consider  the ground state of $\hat{H}_0$ to be of Bijl-Jastrow form 
\beqa
\label{Psi0free}
\Phi_0(x_1,\dots,x_N)=\la x_1,\dots,x_N|\Phi_0\ra=\prod_{i<j}f(x_{ij}),
\eeqa
this is, written as a product  of a single pair correlation function $f(x_{ij})$ over each pair of particles \cite{Sutherland04}.
Here , we denote by $x_{ij}=x_i-x_j$ and note that $f(x_{ij})=\epsilon f(-x_{ij})$ with $\epsilon=1$ for bosons and $\epsilon=-1$ for fermions. (It is actually possibly to consider a more general case of one dimensional anyons \cite{Kundu99,Girardeau06,Batchelor06}.)

In this case, the parent Hamiltonian $\hat{H}_0$ involves exclusively two-body and three-body interactions
\beqa
\label{CalMarH}
\hat{H}_0=-\frac{\hbar^2}{2m}\sum_{i=1}^{N}\frac{\partial^2}{\partial x_i^2} +V_2+V_3,
\eeqa
where the two-body and three-body potentials are given by
\beqa
\label{V2Eq}
V_2&=&\frac{\hbar^2}{m}  \sum_{ i < j} \frac{f''(x_{ij})}{f(x_{ij})},\\
\label{V3Eq}
V_3&=&\frac{\hbar^2}{m}\sum_{i<j<k}\left[\frac{f'(x_{ij})f'(x_{ik})}{f(x_{ij})f(x_{ik})}
-\frac{f'(x_{ij})f'(x_{jk})}{f(x_{ij})f(x_{jk})}+\frac{f'(x_{ik})f'(x_{jk})}{f(x_{ik})f(x_{jk})}
\right].\nonumber\\
\eeqa
Here, $f'$ and $f''$ denote the first and second spatial derivatives of $f$, respectively.
This result  is similar in spirit to that found by Calogero and Marchioro in three-dimensional systems \cite{CalogeroMarchioro75}. However, by focusing on  one-spatial dimension, both $V_2$ and $V_3$ take a simpler form.  Under periodic boundary conditions, the family of Hamiltonians $\hat{H}_0$ that involve exclusively two-body interactions has been studied in detail  \cite{Sutherland04}.  We shall focus instead on systems in the real line, of relevance to the description of ultracold gases in tight-waveguides. As we shall see, $V_3$ effectively vanishes in a number of relevant examples. 

Using (\ref{Psi0free})-(\ref{V3Eq}) we note that the energy eigenvalue of $\Phi_0$ is $E_0=0$, e.g., $\hat{H}_0\Phi_0=0$.
In what follows, we consider the ground state in the presence of a harmonic trap  to be of the form
\beqa
\label{Psi0trap}
\Psi_0
=\exp\left(-\frac{m\om}{2\hbar}\sum_{i=1}^Nx_i^2\right)\prod_{i<j}f(x_{ij}).
\eeqa
By explicit computation, one then finds
\beqa
e^{-\frac{m\om}{2\hbar}\sum_{i=1}^Nx_i^2}\sum^N_{i=1}x_i \partial_{x_i} \prod_{i<j}f(x_{ij})=\sum^N_{i<j}x_{ij}\frac{f'(x_{ij})}{f(x_{ij})}\Psi_0,
\eeqa
which yields the following many-body Schr\"odinger equation
\beqa
\label{trapH0}
\hat{H}\Psi_0=\left[\hat{H}_0+\sum_{i=1}^N\frac{1}{2}m\om^2x_i^2+V_{2{\rm L}}\right]\Psi_0=\frac{N\hbar\om}{2}\Psi_0,
\eeqa
where we have identified the long-range two-body potential 
\beqa
\label{V2long}
V_{2{\rm L}}=-\hbar\om\sum^N_{i<j}x_{ij}\frac{f'(x_{ij})}{f(x_{ij})}.
\eeqa
Equations (\ref{trapH0})-(\ref{V2long}), with $\hat{H}_0$ given by Eqs. (\ref{CalMarH})-(\ref{V3Eq}), provide the complete  family of parent Hamiltonians whose exact ground state  is of Bijl-Jastrow form and given by (\ref{PsiPhiTrap}), describing harmonically-trapped quantum particles subject to two-body and possibly three-body interactions. This is our main result, that we next apply to derive a number of exact many-body solutions, with a focus on cases in which $V_3$ is constant, and thus simply contribute to the energy eigenvalue $\mathcal{E}_0$ of $\Psi_0$. 

{\it Calogero-Sutherland gas.---}
Let us first consider the rational Calogero-Sutherland model \cite{Calogero71,Sutherland71} describing trapped bosonic particles with inverse-square interactions. 
This model describes free bosons in one-dimension for $\lambda=0$ and hard-core bosons in the Tonks-Girardeau regime for $\lambda=1$ \cite{Girardeau60}, a regime experimentally explored in \cite{Kinoshita04,Paredes04}. For arbitrary $\lambda$, the Calogero-Sutherland model can be considered as an ideal gas of particles obeying generalized exclusion statistics \cite{Haldane91,Wu94}, as shown by Murthy and Shankar \cite{MurthyShankar94}. 
In the absence of the trapping potential,  $\hat{H}_0$ describes the Calogero-Moser model,  with ground-state wavefunction $\Phi_0=\prod_{i<j}|x_{ij}|^\lambda$ and energy eigenvalue $E_0=0$. Indeed, for $f(x_{ij})=|x_{ij}|^\lambda$, equation (\ref{V2Eq}) reduces to $V_2=\frac{\hbar^2}{m}\lambda(\lambda-1)/|x_{ij}|^2$ while the three-body term  (\ref{V3Eq}) identically vanishes $V_3=0$.

In the presence of the trap, we consider
$\Psi_0=\exp(-\frac{m\om}{2\hbar}\sum_{i=1}^Nx_i^2)\prod_{i<j}|x_{ij}|^\lambda$.
In this case, the long-range two-body potential reads
\beqa
V_{2{\rm L}}=-\hbar\om\lambda \frac{N(N-1)}{2},
\eeqa
and is thus a constant, resembling a mean-field energy contribution to the ground-state energy. As a result, $\Psi_0$ has energy $\mathcal{E}_0=\frac{\hbar\om}{2}N[1+\lambda(N-1)]$ which reproduces precisely the known expression for the ground-state energy for the (rational) Calogero-Sutherland gas in a harmonic trap \cite{Sutherland71}. As a particular case, for $\lambda=1$,  $\Psi_0$ and $\mathcal{E}_0$ match the corresponding results for the harmonically-trapped  Tonks-Girardeau gas \cite{GWT01}. 

{\it Lieb-Liniger and Lieb-Liniger-Coulomb  gases.---}
The Lieb-Liniger model  with contact-interactions described by a Dirac delta function \cite{LL63,L63} occupies a unique status being both Bethe-ansatz solvable and directly relevant to the description of ultracold gases in tight-waveguides \cite{Olshanii98}. In homogeneous space, the model supports both bright \cite{McGuire64} and dark \cite{Ishikawa80,Deguchi1,Deguchi2} quantum many-body solitons.
Let us consider the wavefunction $\Phi_0=\exp(g\sum_{i<j}|x_{ij}|)$ where $g$ is a coupling constant. This choice is motivated by the fact that for $g<0$, $\Phi_0$ describes a McGuire bright quantum many-body soliton which is an energy eigenstate \cite{McGuire64}. 
We note that  the logarithmic spatial derivative of the pair function $f(x_{ij})=\exp(g|x_{ij}|)$ is given in terms of the sign function $f'(x_{ij})/f(x_{ij})=g{\rm sgn}(x_{ij})$.  Writing the latter in terms of the Heaviside step function $\Theta(x)$ as ${\rm sgn}(x)=2\Theta(x)-1$, and noting that $\frac{d}{dx}\Theta(x)=\delta(x)$ \cite{Bracewell00}, it follows that
$\frac{d}{dx}{\rm sgn}(x)=2\delta(x)$. As a result, the two-body contribution $V_2$ is consistent with  Lieb-Liniger contact interactions.  Specifically,
\beqa
V_2=\frac{\hbar^2}{m}2g\sum_{i<j}\delta(x_{ij})+\frac{g^2\hbar^2}{m}\frac{N(N-1)}{2}, 
\eeqa
which  is precisely the contact pseudopotential that describes $s$-wave scattering, plus an additional constant. We further note  the three-body potential reduces to 
$V_3=\frac{\hbar^2}{m}g^2\frac{N(N-1)(N-2)}{6}$. 
The constant contribution of $V_2$  and $V_3$ add up  precisely to (minus) the energy of the McGuire solution in free space, i.e., 
\beqa
\label{McGuireE0}
E_0=-\frac{g^2\hbar^2}{m}\frac{N(N^2-1)}{6}.
\eeqa 
 We look for the Hamiltonian with ground state 
\beqa
\Psi_0=\exp\left(-\frac{m\om}{2\hbar}\sum_{i=1}^Nx_i^2+g\sum_{i<j}|x_{ij}|\right),
\eeqa
 which  represents a McGuire soliton embedded in a harmonic trap.
In this case, the long range contribution becomes
\beqa
V_{2{\rm L}}=-\hbar\om g\sum_{i<j}|x_{ij}|,
\eeqa
and the Sch\"odinger equation in the presence of the trap reads
\beqa 
\label{LLC}
& & \left[\sum_{i=1}^{N}\left(-\frac{\hbar^2}{2m}\frac{\partial^2}{\partial x_i^2} + \frac{1}{2}m\om^2x_i^2\right)+g \sum_{ i < j} \left(\frac{\hbar^2}{m}2\delta (x_{ij})-\frac{m\om }{\hbar} |x_{ij}|\right)\right]\Psi_0\nonumber\\
& & =\left(E_0+\frac{N\hbar\om}{2}\right)\Psi_0.
\eeqa
This Hamiltonian is precisely the Lieb-Liniger-Coulomb model recently introduced in \cite{BeauPittman20} using Polychronakos formalism \cite{Polychronakos92}. The model describes one dimensional bosons subject to contact interactions as well as a long-range pairwise contribution. The latter can be thought of accounting a one-dimensional repulsive Coulomb interaction for $g>0$ or an attractive gravitational potential for $g<0$  \cite{Muriel76}. Interestingly, the coupling strengths of the  contact and long-range  interaction terms are not independent, and in the limit of vanishing trapping frequency the long-range contribution is absent and one recovers the Lieb-Liniger model in free-space. This is the general feature of the family of models (\ref{trapH0}).

{\it Gaussian pair function.---}
As a generalization of the McGuire bright soliton solution we  consider $f(x_{ij})=\exp(g|x_{ij}|^2)$, allowing for arbitrary sign of $g$. The  case with $g<0$ describes a Gaussian pair function that  is commonly used in Monte Carlo calculations   for quantum fluids. The corresponding Bijl-Jastrow many-body wavefunction is the ground state of the Hamiltonian $\hat{H}_0$ in Eq. (\ref{CalMarH}) in which the two-body term is
\beqa
V_2=\frac{\hbar^2}{m}gN(N-1) +\frac{\hbar^2}{m}4g^2
\sum_{i<j}|x_{ij}|^2.
\eeqa 
For the Gaussian pair function, the  three-body contribution admits an expression compatible with a two-body potential as 
\beqa
V_3=\frac{\hbar^2}{m}g^2(2N-4)\sum_{i<j}|x_{ij}|^2,
\eeqa 
and thus,
$V_2+V_3=-E_0+\frac{\hbar^2}{m}2Ng^2\sum_{i<j}|x_{ij}|^2$. Here, 
the ground-state energy in free space equals $E_0=-gN(N-1)\frac{\hbar^2}{m}$, where we note the linear and quadratic scaling with $g$ and $N$, respectively.  The full interaction potential is  described by pairwise  attractive quadratic terms of the interparticle distance. Upon embedding in a harmonic trap,  the long-range contribution reads
\beqa
V_{2{\rm L}}=-2\hbar\om g\sum^N_{i<j}|x_{ij}|^2,
\eeqa
and the ground-state wavefunction $\Psi_0=\exp\left(-\frac{m\om}{2\hbar}\sum_{i=1}^Nx_i^2+g\sum_{i<j}|x_{ij}|^2\right)$ has  energy 
$\mathcal{E}_0=\frac{N\hbar\om}{2}-gN(N-1)\frac{\hbar^2}{m}$. 
By contrast, when the pair function $f(x_{ij})$ is chosen as a hyper-Gaussian, $f(x_{ij})=\exp(g|x_{ij}|^n)$, the parent Hamiltonian $H_0$ in the homogeneous case involves non-trivial three-body interactions of the general form in Eq. (\ref{V3Eq}).

{\it Hyperbolic pair functions.---}
Another relevant system is the one associated with the following choice of the pair correlation function
$f(x_{ij})=\sinh(x_{ij}/\ell)^\lambda$, which describes bosons for even $\lambda$. We note that $f(x_{ij})\sim \exp(\lambda |x_{ij}|/\ell)$ for large $x_{ij}$, so one can expect a similarity with the Lieb-Liniger models discussed. However, the hyperbolic  $f(x_{ij})$ includes a hard-core constraint as $f(x_{ij}=0)=0$ and grows faster with the interparticle distance than the pair function in the Calogero-Sutherland model, $f(x_{ij})=|x_{ij}|^\lambda$.
In addition, the wavefunction at contact $x_{ij}=0$ is continuous in this model.
In this case,
\beqa
V_2&=&\frac{\lambda\hbar^2N(N-1)}{2m\ell^2}+\frac{\hbar^2}{m\ell^2}\sum_{i<j}\frac{\lambda(\lambda-1)}{{\rm sinh}^2(x_{ij}/\ell)},\\
V_3&=&\frac{\lambda^2\hbar^2N(N-1)(N-2)}{6m\ell^2}. \eeqa
The pairwise hyperbolic potential diverges at the origin, imposing the hard-core constraint between particles. At low-density it decays exponentially with the interparticle distance, as the one in the Toda lattice \cite{Toda67,Sutherland04}.
For the ground state in free space $\Phi_0=\prod_{i<j}\sinh(x_{ij}/\ell)^\lambda$, the energy is 
$E_0=-\lambda^2\frac{\hbar^2}{6m\ell^2}N(N^2-1)$, which resembles that of  a McGuire solution in free space; see  (\ref{McGuireE0}).
In the presence of a harmonic trap, the long-range contribution reads
\beqa
V_{2{\rm L}}=\hbar\lambda \om  \sum^N_{i<j}\frac{x_{ij}}{\ell}{\rm coth}\left(\frac{x_{ij}}{\ell}\right).
\eeqa
We note that $x{\rm coth}x\sim |x|$ for large $x$, as in the Lieb-Liniger-Coulomb model (\ref{LLC}), but being attractive, it behaves as a one-dimensional  gravitational potential \cite{Muriel76}. In addition,  it is also continuous and effectively harmonic near the origin, i.e.,  $x{\rm coth}x=1+x^2/3+\mathcal{O}(x^4)$. The energy of the trapped ground state $\Psi_0=\exp(-\frac{m\om}{2\hbar}\sum_{i=1}^Nx_i^2)\prod_{i<j}\sinh(x_{ij}/\ell)^\lambda$ is $\mathcal{E}_0=\frac{N\hbar\om}{2}-\lambda^2\frac{\hbar^2}{6m\ell^2}N(N^2-1)$.

{\it Anharmonic  external potentials.---}
The previous  examples illustrate the validity of the framework put forward, in which a Hamiltonian of the form (\ref{CalMarH}) with ground state (\ref{Psi0free}) can be used to construct a Hamiltonian (\ref{trapH0}) involving a harmonic trap and generally long-range pairwise interactions, with ground state 
(\ref{Psi0trap}). This setting can be generalized to include anharmonic external potentials.
To this end,  we next consider the ground state of the Hamiltonian with confinement to be described by a state 
\beqa
\label{Psi0Gen}
\Psi_0=\exp\left(\sum_{i=1}^Nv(x_i)\right)\Phi_0,
\eeqa
 where $v(x)$ is an arbitrary function of the coordinates.  In this case, $\Psi_0$ is a solution of
\beqa
\left[\hat{H}_0+\sum_{i=1}^NV(x_i)\right]\Psi_0
+\frac{\hbar^2}{m}e^{\sum_{i=1}^Nv(x_i)}
\sum_{i=1}^Nv'(x_i)\partial_{x_i}
\Phi_0=0,\nonumber\\
\eeqa
where the one-body potential reads
\beqa
\label{Vtrap}
V(x)=\frac{\hbar^2}{2m}\left[v''(x)+v'(x)^2\right].
\eeqa
Assuming $\Phi_0$ to have a Jastrow form as in Eq. (\ref{Psi0free}), it follows that parent Hamiltonian of the generalized ground state (\ref{Psi0Gen}) is 
\beqa
\label{GenTISE}
\left[\hat{H}_0+\sum_{i=1}^NV(x_i)\right]\Psi_0
+\frac{\hbar^2}{m}\sum^N_{i<j}[v'(x_i)-v'(x_j)]\frac{f'(x_{ij})}{f(x_{ij})}
\Psi_0=0,\nonumber\\
\eeqa
with energy eigenvalue $\mathcal{E}_0=0$.
We note that for a specific form of $V(x)$, equation (\ref{Vtrap}) can be integrated to find the function $v(x)$.

{\it Quantum solids.---} 
An approach to characterize many-body states of quantum solids relies on pinning particles on a lattice, 
without relying on a tight-binding description, e.g., as done in the description of solid $^4$He \cite{Cazorla07,Cazorla09}. Motivated by these approach, we next consider the family of Nosanov-Jastrow wavefunctions of the form
\beqa
\label{NJPsi}
\Psi_0=\prod_{i=1}^N\exp\left(-\frac{m\om}{2\hbar}(x_i-x_i^0)^2\right)\Phi_0,
\eeqa
where the Gaussian term localizes the $i$-th particle at a position $x_i^0$ and $\om$ controls the tightness of the confinement. For instance, one can choose $x_i^0=ia$ where $a$ is the lattice spacing. Two-particle correlations are encoded in $\Phi_0$ given by Eq. (\ref{Psi0free}). 
The many-body Schr\"odinger equation (\ref{GenTISE}) reduces to
\beqa
\left[\hat{H}_0+\sum_{i=1}^N\frac{m\om^2}{2}\delta x_i^2
-\hbar\om\sum^N_{i<j}\delta x_{ij}\frac{f'(x_{ij})}{f(x_{ij})}
\right]\Psi_0=\frac{N\hbar\om}{2}\Psi_0,\nonumber\\
\eeqa
where the term in square brackets is the parent Hamiltonian of (\ref{NJPsi}), with $\delta x_{i}=x_i-x_i^0$ and $\delta x_{ij}=\delta x_{i}-\delta x_{j}=x_{ij}-(x_i^0-x_j^0)$. Particles are thus confined by a lattice  of harmonic oscillator wells and anharmonic effects can be included by considering a more general confining potential. We note that $\hat{H}_0$ is not altered and it is thus straightforward to introduce the quantum-solid Hamiltonians associated with the models discussed above. 
As an example,  the  quantum-solid version of the Calogero-Sutherland model is described by the Hamiltonian
\beqa
\hat{H}=
\sum_{i=1}^{N}\left(-\frac{\hbar^2}{2m}\frac{\partial^2}{\partial x_i^2}+ \frac{m\om^2}{2}\delta x_i^2\right)
+\sum^N_{i<j}\left(\frac{\lambda(\lambda-1)}{|x_{ij}|^2}+\hbar\om \lambda\frac{\delta x_{ij}^0}{x_{ij}}\right),\nonumber\\
\eeqa
which includes an  additional Coulomb-like term, in which $\delta x_{ij}^0=x_i^0-x_j^0$ is constant for fixed sites  indices $i$ and $j$. Its ground state reads $\Psi_0=\prod_{i=1}^N\exp(-\frac{m\om}{2\hbar}(x_i-x_i^0)^2)\prod_{i<j}|x_{ij}|^\lambda$ with energy eigenvalue $\mathcal{E}_0=\frac{N\hbar\om}{2}[1+\lambda(N-1)]$.  For $\lambda=1$ one recovers the hard-core bosonic pair correlation function. 
Due to the lattice structure, permutation symmetry is explicitly broken as the $i$-th particle is localized in the $i$-th harmonic well, but it can be restored by explicit symmetrization, after which the the ground-state wavefunction becomes $\Psi_0=[\sum_{P\in S_n}\prod_{i=1}^N\exp(-\frac{m\om}{2\hbar}(x_i-x_{P(i)}^0)^2)]\prod_{i<j}|x_{ij}|^\lambda$, where the sum runs over the $N!$ permutations of the $N$ lattice sites and $S_N$ denotes the symmetric group. In this case, $\Psi_0$ with $\lambda=1$ describes a Tonks-Girardeau gas in a lattice of harmonic wells.

{\it Summary.---} We have introduced an exact mapping between the ground state of a Hamiltonian in free space and the ground state of a dual Hamiltonian in the presence of a one-body trapping potential and additional many-body interactions.  Whenever the homogeneous ground state takes the Bijl-Jastrow form, the dual Hamiltonian  can be expressed in terms of the homogeneous one, supplemented with  the one-body potential and two-body long-range interactions. 
This mapping  can be used by fixing the pair-correlation function entering the Bijl-Jastrow form,  as we have done to find trapped states in systems with inverse-square, contact  and quadratic interactions. As an alternative,  the functional form of the interparticle interactions can be first established, from which the pair function can be determined by integration. In addition, we have also shown how this family of Hamiltonians can be generalized to anharmonic external potentials and the description of quantum solids with ground states of  Nosanov-Jastrow form.
Our results, should be broadly applicable in the quest of  novel beautiful models,  describing (quasi-) solvable and integrable quantum many-body systems. Possible extensions include  systems in higher spatial dimensions  \cite{CalogeroMarchioro75,KhareRay97}, mixtures of multiple species \cite{Ujino98,GirardeauMinguzzi07}, particles with internal-structure (e.g. spinors) \cite{Girardeau04}, and their variants with truncated interaction range \cite{JainKhare99,Auberson01,Basu-Mallick01,Pittman17,Tummuru17}. Excited states in these systems can be explored by established techniques \cite{Sutherland04},  and our results
 should  also find applications in the study of bright and dark trapped quantum many-body solitons.

{\it Acknowledgements.---} It is a pleasure to acknowledge discussions with Gregory Astrakarchik, Mathieu Beau, Francesco Calogero, \'I\~{n}igo L. Egusquiza, Xi-Wen Guan, Avinash Khare 
and Suzanne Pittman.

\bibliography{BAIQS_lib}

\begin{thebibliography}{56}%
\makeatletter
\providecommand \@ifxundefined [1]{%
 \@ifx{#1\undefined}
}%
\providecommand \@ifnum [1]{%
 \ifnum #1\expandafter \@firstoftwo
 \else \expandafter \@secondoftwo
 \fi
}%
\providecommand \@ifx [1]{%
 \ifx #1\expandafter \@firstoftwo
 \else \expandafter \@secondoftwo
 \fi
}%
\providecommand \natexlab [1]{#1}%
\providecommand \enquote  [1]{``#1''}%
\providecommand \bibnamefont  [1]{#1}%
\providecommand \bibfnamefont [1]{#1}%
\providecommand \citenamefont [1]{#1}%
\providecommand \href@noop [0]{\@secondoftwo}%
\providecommand \href [0]{\begingroup \@sanitize@url \@href}%
\providecommand \@href[1]{\@@startlink{#1}\@@href}%
\providecommand \@@href[1]{\endgroup#1\@@endlink}%
\providecommand \@sanitize@url [0]{\catcode `\\12\catcode `\$12\catcode
  `\&12\catcode `\#12\catcode `\^12\catcode `\_12\catcode `\%12\relax}%
\providecommand \@@startlink[1]{}%
\providecommand \@@endlink[0]{}%
\providecommand \url  [0]{\begingroup\@sanitize@url \@url }%
\providecommand \@url [1]{\endgroup\@href {#1}{\urlprefix }}%
\providecommand \urlprefix  [0]{URL }%
\providecommand \Eprint [0]{\href }%
\providecommand \doibase [0]{http://dx.doi.org/}%
\providecommand \selectlanguage [0]{\@gobble}%
\providecommand \bibinfo  [0]{\@secondoftwo}%
\providecommand \bibfield  [0]{\@secondoftwo}%
\providecommand \translation [1]{[#1]}%
\providecommand \BibitemOpen [0]{}%
\providecommand \bibitemStop [0]{}%
\providecommand \bibitemNoStop [0]{.\EOS\space}%
\providecommand \EOS [0]{\spacefactor3000\relax}%
\providecommand \BibitemShut  [1]{\csname bibitem#1\endcsname}%
\let\auto@bib@innerbib\@empty
\bibitem [{\citenamefont {Sutherland}(2004)}]{Sutherland04}%
  \BibitemOpen
  \bibfield  {author} {\bibinfo {author} {\bibfnamefont {B.}~\bibnamefont
  {Sutherland}},\ }\href {\doibase 10.1142/5552} {\emph {\bibinfo {title}
  {Beautiful Models}}}\ (\bibinfo  {publisher} {World Scientific},\ \bibinfo
  {year} {2004})\BibitemShut {NoStop}%
\bibitem [{\citenamefont {Korepin}\ \emph {et~al.}(1997)\citenamefont
  {Korepin}, \citenamefont {Bogoliubov},\ and\ \citenamefont
  {Izergin}}]{KBI97}%
  \BibitemOpen
  \bibfield  {author} {\bibinfo {author} {\bibfnamefont {V.~E.}\ \bibnamefont
  {Korepin}}, \bibinfo {author} {\bibfnamefont {N.~M.}\ \bibnamefont
  {Bogoliubov}}, \ and\ \bibinfo {author} {\bibfnamefont {A.~G.}\ \bibnamefont
  {Izergin}},\ }\href@noop {} {\emph {\bibinfo {title} {Quantum Inverse
  Scattering Method and Correlation Functions}}}\ (\bibinfo  {publisher}
  {Cambridge, Cambridge},\ \bibinfo {year} {1997})\BibitemShut {NoStop}%
\bibitem [{\citenamefont {Takahashi}(1999)}]{Takahashi99}%
  \BibitemOpen
  \bibfield  {author} {\bibinfo {author} {\bibfnamefont {M.}~\bibnamefont
  {Takahashi}},\ }\href@noop {} {\emph {\bibinfo {title} {Thermodynamics of
  One-Dimensional Solvable Models}}}\ (\bibinfo  {publisher} {Cambridge,
  Cambridge},\ \bibinfo {year} {1999})\BibitemShut {NoStop}%
\bibitem [{\citenamefont {Gaudin}(2014)}]{Gaudin14}%
  \BibitemOpen
  \bibfield  {author} {\bibinfo {author} {\bibfnamefont {M.}~\bibnamefont
  {Gaudin}},\ }\href {\doibase 10.1017/CBO9781107053885} {\emph {\bibinfo
  {title} {The Bethe Wavefunction}}},\ edited by\ \bibinfo {editor}
  {\bibfnamefont {J.-S.}\ \bibnamefont {Caux}}\ (\bibinfo  {publisher}
  {Cambridge University Press},\ \bibinfo {year} {2014})\BibitemShut {NoStop}%
\bibitem [{\citenamefont {Gaudin}(1971)}]{Gaudin71}%
  \BibitemOpen
  \bibfield  {author} {\bibinfo {author} {\bibfnamefont {M.}~\bibnamefont
  {Gaudin}},\ }\href {\doibase 10.1103/PhysRevA.4.386} {\bibfield  {journal}
  {\bibinfo  {journal} {Phys. Rev. A}\ }\textbf {\bibinfo {volume} {4}},\
  \bibinfo {pages} {386} (\bibinfo {year} {1971})}\BibitemShut {NoStop}%
\bibitem [{\citenamefont {Batchelor}\ \emph {et~al.}(2006)\citenamefont
  {Batchelor}, \citenamefont {Guan},\ and\ \citenamefont
  {Oelkers}}]{Batchelor06}%
  \BibitemOpen
  \bibfield  {author} {\bibinfo {author} {\bibfnamefont {M.~T.}\ \bibnamefont
  {Batchelor}}, \bibinfo {author} {\bibfnamefont {X.-W.}\ \bibnamefont {Guan}},
  \ and\ \bibinfo {author} {\bibfnamefont {N.}~\bibnamefont {Oelkers}},\ }\href
  {\doibase 10.1103/PhysRevLett.96.210402} {\bibfield  {journal} {\bibinfo
  {journal} {Phys. Rev. Lett.}\ }\textbf {\bibinfo {volume} {96}},\ \bibinfo
  {pages} {210402} (\bibinfo {year} {2006})}\BibitemShut {NoStop}%
\bibitem [{\citenamefont {Calogero}(1971)}]{Calogero71}%
  \BibitemOpen
  \bibfield  {author} {\bibinfo {author} {\bibfnamefont {F.}~\bibnamefont
  {Calogero}},\ }\href {\doibase 10.1063/1.1665604} {\bibfield  {journal}
  {\bibinfo  {journal} {Journal of Mathematical Physics}\ }\textbf {\bibinfo
  {volume} {12}},\ \bibinfo {pages} {419} (\bibinfo {year} {1971})}\BibitemShut
  {NoStop}%
\bibitem [{\citenamefont {Sutherland}(1971)}]{Sutherland71}%
  \BibitemOpen
  \bibfield  {author} {\bibinfo {author} {\bibfnamefont {B.}~\bibnamefont
  {Sutherland}},\ }\href {\doibase 10.1063/1.1665584} {\bibfield  {journal}
  {\bibinfo  {journal} {Journal of Mathematical Physics}\ }\textbf {\bibinfo
  {volume} {12}},\ \bibinfo {pages} {246} (\bibinfo {year} {1971})}\BibitemShut
  {NoStop}%
\bibitem [{\citenamefont {Girardeau}(1960)}]{Girardeau60}%
  \BibitemOpen
  \bibfield  {author} {\bibinfo {author} {\bibfnamefont {M.}~\bibnamefont
  {Girardeau}},\ }\href {\doibase 10.1063/1.1703687} {\bibfield  {journal}
  {\bibinfo  {journal} {Journal of Mathematical Physics}\ }\textbf {\bibinfo
  {volume} {1}},\ \bibinfo {pages} {516} (\bibinfo {year} {1960})}\BibitemShut
  {NoStop}%
\bibitem [{\citenamefont {Girardeau}\ \emph {et~al.}(2001)\citenamefont
  {Girardeau}, \citenamefont {Wright},\ and\ \citenamefont {Triscari}}]{GWT01}%
  \BibitemOpen
  \bibfield  {author} {\bibinfo {author} {\bibfnamefont {M.~D.}\ \bibnamefont
  {Girardeau}}, \bibinfo {author} {\bibfnamefont {E.~M.}\ \bibnamefont
  {Wright}}, \ and\ \bibinfo {author} {\bibfnamefont {J.~M.}\ \bibnamefont
  {Triscari}},\ }\href {\doibase 10.1103/PhysRevA.63.033601} {\bibfield
  {journal} {\bibinfo  {journal} {Phys. Rev. A}\ }\textbf {\bibinfo {volume}
  {63}},\ \bibinfo {pages} {033601} (\bibinfo {year} {2001})}\BibitemShut
  {NoStop}%
\bibitem [{\citenamefont {Girardeau}(2006)}]{Girardeau06}%
  \BibitemOpen
  \bibfield  {author} {\bibinfo {author} {\bibfnamefont {M.~D.}\ \bibnamefont
  {Girardeau}},\ }\href {\doibase 10.1103/PhysRevLett.97.100402} {\bibfield
  {journal} {\bibinfo  {journal} {Phys. Rev. Lett.}\ }\textbf {\bibinfo
  {volume} {97}},\ \bibinfo {pages} {100402} (\bibinfo {year}
  {2006})}\BibitemShut {NoStop}%
\bibitem [{\citenamefont {del Campo}(2008)}]{delcampo08}%
  \BibitemOpen
  \bibfield  {author} {\bibinfo {author} {\bibfnamefont {A.}~\bibnamefont {del
  Campo}},\ }\href {\doibase 10.1103/PhysRevA.78.045602} {\bibfield  {journal}
  {\bibinfo  {journal} {Phys. Rev. A}\ }\textbf {\bibinfo {volume} {78}},\
  \bibinfo {pages} {045602} (\bibinfo {year} {2008})}\BibitemShut {NoStop}%
\bibitem [{\citenamefont {Ujino}\ \emph {et~al.}(1998)\citenamefont {Ujino},
  \citenamefont {Nishino},\ and\ \citenamefont {Wadati}}]{Ujino98}%
  \BibitemOpen
  \bibfield  {author} {\bibinfo {author} {\bibfnamefont {H.}~\bibnamefont
  {Ujino}}, \bibinfo {author} {\bibfnamefont {A.}~\bibnamefont {Nishino}}, \
  and\ \bibinfo {author} {\bibfnamefont {M.}~\bibnamefont {Wadati}},\ }\href
  {\doibase https://doi.org/10.1016/S0375-9601(98)00769-5} {\bibfield
  {journal} {\bibinfo  {journal} {Physics Letters A}\ }\textbf {\bibinfo
  {volume} {249}},\ \bibinfo {pages} {459 } (\bibinfo {year}
  {1998})}\BibitemShut {NoStop}%
\bibitem [{\citenamefont {Girardeau}\ and\ \citenamefont
  {Minguzzi}(2007)}]{GirardeauMinguzzi07}%
  \BibitemOpen
  \bibfield  {author} {\bibinfo {author} {\bibfnamefont {M.~D.}\ \bibnamefont
  {Girardeau}}\ and\ \bibinfo {author} {\bibfnamefont {A.}~\bibnamefont
  {Minguzzi}},\ }\href {\doibase 10.1103/PhysRevLett.99.230402} {\bibfield
  {journal} {\bibinfo  {journal} {Phys. Rev. Lett.}\ }\textbf {\bibinfo
  {volume} {99}},\ \bibinfo {pages} {230402} (\bibinfo {year}
  {2007})}\BibitemShut {NoStop}%
\bibitem [{\citenamefont {Kundu}(2009)}]{Kundu09}%
  \BibitemOpen
  \bibfield  {author} {\bibinfo {author} {\bibfnamefont {A.}~\bibnamefont
  {Kundu}},\ }\href {\doibase 10.1103/PhysRevE.79.015601} {\bibfield  {journal}
  {\bibinfo  {journal} {Phys. Rev. E}\ }\textbf {\bibinfo {volume} {79}},\
  \bibinfo {pages} {015601} (\bibinfo {year} {2009})}\BibitemShut {NoStop}%
\bibitem [{\citenamefont {Giorgini}\ \emph {et~al.}(2008)\citenamefont
  {Giorgini}, \citenamefont {Pitaevskii},\ and\ \citenamefont
  {Stringari}}]{Giorgini08}%
  \BibitemOpen
  \bibfield  {author} {\bibinfo {author} {\bibfnamefont {S.}~\bibnamefont
  {Giorgini}}, \bibinfo {author} {\bibfnamefont {L.~P.}\ \bibnamefont
  {Pitaevskii}}, \ and\ \bibinfo {author} {\bibfnamefont {S.}~\bibnamefont
  {Stringari}},\ }\href {\doibase 10.1103/RevModPhys.80.1215} {\bibfield
  {journal} {\bibinfo  {journal} {Rev. Mod. Phys.}\ }\textbf {\bibinfo {volume}
  {80}},\ \bibinfo {pages} {1215} (\bibinfo {year} {2008})}\BibitemShut
  {NoStop}%
\bibitem [{\citenamefont {Tan}(2004)}]{tan2004short}%
  \BibitemOpen
  \bibfield  {author} {\bibinfo {author} {\bibfnamefont {S.}~\bibnamefont
  {Tan}},\ }\href@noop {} {\enquote {\bibinfo {title} {Short range scaling laws
  of quantum gases with contact interactions},}\ } (\bibinfo {year} {2004}),\
  \Eprint {http://arxiv.org/abs/cond-mat/0412764} {arXiv:cond-mat/0412764
  [cond-mat.stat-mech]} \BibitemShut {NoStop}%
\bibitem [{\citenamefont {Werner}\ and\ \citenamefont
  {Castin}(2006)}]{WernerCastin06}%
  \BibitemOpen
  \bibfield  {author} {\bibinfo {author} {\bibfnamefont {F.}~\bibnamefont
  {Werner}}\ and\ \bibinfo {author} {\bibfnamefont {Y.}~\bibnamefont
  {Castin}},\ }\href {\doibase 10.1103/PhysRevA.74.053604} {\bibfield
  {journal} {\bibinfo  {journal} {Phys. Rev. A}\ }\textbf {\bibinfo {volume}
  {74}},\ \bibinfo {pages} {053604} (\bibinfo {year} {2006})}\BibitemShut
  {NoStop}%
\bibitem [{\citenamefont {Castin}\ and\ \citenamefont
  {Werner}(2012)}]{Castin12}%
  \BibitemOpen
  \bibfield  {author} {\bibinfo {author} {\bibfnamefont {Y.}~\bibnamefont
  {Castin}}\ and\ \bibinfo {author} {\bibfnamefont {F.}~\bibnamefont
  {Werner}},\ }\enquote {\bibinfo {title} {The unitary gas and its symmetry
  properties},}\ in\ \href {\doibase 10.1007/978-3-642-21978-8_5} {\emph
  {\bibinfo {booktitle} {The BCS-BEC Crossover and the Unitary Fermi Gas}}},\
  \bibinfo {editor} {edited by\ \bibinfo {editor} {\bibfnamefont
  {W.}~\bibnamefont {Zwerger}}}\ (\bibinfo  {publisher} {Springer Berlin
  Heidelberg},\ \bibinfo {address} {Berlin, Heidelberg},\ \bibinfo {year}
  {2012})\ pp.\ \bibinfo {pages} {127--191}\BibitemShut {NoStop}%
\bibitem [{\citenamefont {Kawakami}(1993)}]{Kawakami93}%
  \BibitemOpen
  \bibfield  {author} {\bibinfo {author} {\bibfnamefont {N.}~\bibnamefont
  {Kawakami}},\ }\href {\doibase 10.1143/JPSJ.62.4163} {\bibfield  {journal}
  {\bibinfo  {journal} {Journal of the Physical Society of Japan}\ }\textbf
  {\bibinfo {volume} {62}},\ \bibinfo {pages} {4163} (\bibinfo {year}
  {1993})}\BibitemShut {NoStop}%
\bibitem [{\citenamefont {Gurappa}\ \emph {et~al.}(1998)\citenamefont
  {Gurappa}, \citenamefont {Khare},\ and\ \citenamefont
  {Panigrahi}}]{Gurappa98}%
  \BibitemOpen
  \bibfield  {author} {\bibinfo {author} {\bibfnamefont {N.}~\bibnamefont
  {Gurappa}}, \bibinfo {author} {\bibfnamefont {A.}~\bibnamefont {Khare}}, \
  and\ \bibinfo {author} {\bibfnamefont {P.~K.}\ \bibnamefont {Panigrahi}},\
  }\href {\doibase https://doi.org/10.1016/S0375-9601(98)00471-X} {\bibfield
  {journal} {\bibinfo  {journal} {Physics Letters A}\ }\textbf {\bibinfo
  {volume} {244}},\ \bibinfo {pages} {467 } (\bibinfo {year}
  {1998})}\BibitemShut {NoStop}%
\bibitem [{\citenamefont {Gurappa}\ and\ \citenamefont
  {Panigrahi}(1999)}]{Gurappa99}%
  \BibitemOpen
  \bibfield  {author} {\bibinfo {author} {\bibfnamefont {N.}~\bibnamefont
  {Gurappa}}\ and\ \bibinfo {author} {\bibfnamefont {P.~K.}\ \bibnamefont
  {Panigrahi}},\ }\href {\doibase 10.1103/PhysRevB.59.R2490} {\bibfield
  {journal} {\bibinfo  {journal} {Phys. Rev. B}\ }\textbf {\bibinfo {volume}
  {59}},\ \bibinfo {pages} {R2490} (\bibinfo {year} {1999})}\BibitemShut
  {NoStop}%
\bibitem [{\citenamefont {Vacek}\ \emph {et~al.}(1994)\citenamefont {Vacek},
  \citenamefont {Okiji},\ and\ \citenamefont {Kawakami}}]{Vacek94}%
  \BibitemOpen
  \bibfield  {author} {\bibinfo {author} {\bibfnamefont {K.}~\bibnamefont
  {Vacek}}, \bibinfo {author} {\bibfnamefont {A.}~\bibnamefont {Okiji}}, \ and\
  \bibinfo {author} {\bibfnamefont {N.}~\bibnamefont {Kawakami}},\ }\href
  {\doibase 10.1088/0305-4470/27/7/002} {\bibfield  {journal} {\bibinfo
  {journal} {Journal of Physics A: Mathematical and General}\ }\textbf
  {\bibinfo {volume} {27}},\ \bibinfo {pages} {L201} (\bibinfo {year}
  {1994})}\BibitemShut {NoStop}%
\bibitem [{\citenamefont {Kawakami}(1994)}]{Kawakami94}%
  \BibitemOpen
  \bibfield  {author} {\bibinfo {author} {\bibfnamefont {N.}~\bibnamefont
  {Kawakami}},\ }\href {\doibase 10.1143/ptp/91.2.189} {\bibfield  {journal}
  {\bibinfo  {journal} {Progress of Theoretical Physics}\ }\textbf {\bibinfo
  {volume} {91}},\ \bibinfo {pages} {189} (\bibinfo {year} {1994})}\BibitemShut
  {NoStop}%
\bibitem [{\citenamefont {Sogo}(1994)}]{Sogo94}%
  \BibitemOpen
  \bibfield  {author} {\bibinfo {author} {\bibfnamefont {K.}~\bibnamefont
  {Sogo}},\ }\href {\doibase 10.1143/JPSJ.63.879} {\bibfield  {journal}
  {\bibinfo  {journal} {Journal of the Physical Society of Japan}\ }\textbf
  {\bibinfo {volume} {63}},\ \bibinfo {pages} {879} (\bibinfo {year}
  {1994})}\BibitemShut {NoStop}%
\bibitem [{\citenamefont {Sogo}(1996)}]{Sogo96}%
  \BibitemOpen
  \bibfield  {author} {\bibinfo {author} {\bibfnamefont {K.}~\bibnamefont
  {Sogo}},\ }\href {\doibase 10.1143/JPSJ.65.3097} {\bibfield  {journal}
  {\bibinfo  {journal} {Journal of the Physical Society of Japan}\ }\textbf
  {\bibinfo {volume} {65}},\ \bibinfo {pages} {3097} (\bibinfo {year}
  {1996})}\BibitemShut {NoStop}%
\bibitem [{\citenamefont {Jain}\ and\ \citenamefont
  {Khare}(1999)}]{JainKhare99}%
  \BibitemOpen
  \bibfield  {author} {\bibinfo {author} {\bibfnamefont {S.~R.}\ \bibnamefont
  {Jain}}\ and\ \bibinfo {author} {\bibfnamefont {A.}~\bibnamefont {Khare}},\
  }\href {\doibase https://doi.org/10.1016/S0375-9601(99)00637-4} {\bibfield
  {journal} {\bibinfo  {journal} {Physics Letters A}\ }\textbf {\bibinfo
  {volume} {262}},\ \bibinfo {pages} {35 } (\bibinfo {year}
  {1999})}\BibitemShut {NoStop}%
\bibitem [{\citenamefont {Auberson}\ \emph {et~al.}(2001)\citenamefont
  {Auberson}, \citenamefont {Jain},\ and\ \citenamefont {Khare}}]{Auberson01}%
  \BibitemOpen
  \bibfield  {author} {\bibinfo {author} {\bibfnamefont {G.}~\bibnamefont
  {Auberson}}, \bibinfo {author} {\bibfnamefont {S.~R.}\ \bibnamefont {Jain}},
  \ and\ \bibinfo {author} {\bibfnamefont {A.}~\bibnamefont {Khare}},\ }\href
  {\doibase 10.1088/0305-4470/34/4/302} {\bibfield  {journal} {\bibinfo
  {journal} {Journal of Physics A: Mathematical and General}\ }\textbf
  {\bibinfo {volume} {34}},\ \bibinfo {pages} {695} (\bibinfo {year}
  {2001})}\BibitemShut {NoStop}%
\bibitem [{\citenamefont {Basu-Mallick}\ and\ \citenamefont
  {Kundu}(2001)}]{Basu-Mallick01}%
  \BibitemOpen
  \bibfield  {author} {\bibinfo {author} {\bibfnamefont {B.}~\bibnamefont
  {Basu-Mallick}}\ and\ \bibinfo {author} {\bibfnamefont {A.}~\bibnamefont
  {Kundu}},\ }\href {\doibase https://doi.org/10.1016/S0375-9601(00)00816-1}
  {\bibfield  {journal} {\bibinfo  {journal} {Physics Letters A}\ }\textbf
  {\bibinfo {volume} {279}},\ \bibinfo {pages} {29 } (\bibinfo {year}
  {2001})}\BibitemShut {NoStop}%
\bibitem [{\citenamefont {Ezung}\ \emph {et~al.}(2005)\citenamefont {Ezung},
  \citenamefont {Gurappa}, \citenamefont {Khare},\ and\ \citenamefont
  {Panigrahi}}]{Ezung05}%
  \BibitemOpen
  \bibfield  {author} {\bibinfo {author} {\bibfnamefont {M.}~\bibnamefont
  {Ezung}}, \bibinfo {author} {\bibfnamefont {N.}~\bibnamefont {Gurappa}},
  \bibinfo {author} {\bibfnamefont {A.}~\bibnamefont {Khare}}, \ and\ \bibinfo
  {author} {\bibfnamefont {P.~K.}\ \bibnamefont {Panigrahi}},\ }\href {\doibase
  10.1103/PhysRevB.71.125121} {\bibfield  {journal} {\bibinfo  {journal} {Phys.
  Rev. B}\ }\textbf {\bibinfo {volume} {71}},\ \bibinfo {pages} {125121}
  (\bibinfo {year} {2005})}\BibitemShut {NoStop}%
\bibitem [{\citenamefont {Enciso}\ \emph {et~al.}(2005)\citenamefont {Enciso},
  \citenamefont {Finkel}, \citenamefont {Gonz\'alez-L\'opez},\ and\
  \citenamefont {Rodr\'iguez}}]{Enciso05}%
  \BibitemOpen
  \bibfield  {author} {\bibinfo {author} {\bibfnamefont {A.}~\bibnamefont
  {Enciso}}, \bibinfo {author} {\bibfnamefont {F.}~\bibnamefont {Finkel}},
  \bibinfo {author} {\bibfnamefont {A.}~\bibnamefont {Gonz\'alez-L\'opez}}, \
  and\ \bibinfo {author} {\bibfnamefont {M.}~\bibnamefont {Rodr\'iguez}},\
  }\href {\doibase https://doi.org/10.1016/j.physletb.2004.11.031} {\bibfield
  {journal} {\bibinfo  {journal} {Physics Letters B}\ }\textbf {\bibinfo
  {volume} {605}},\ \bibinfo {pages} {214 } (\bibinfo {year}
  {2005})}\BibitemShut {NoStop}%
\bibitem [{\citenamefont {Pittman}\ \emph {et~al.}(2017)\citenamefont
  {Pittman}, \citenamefont {Beau}, \citenamefont {Olshanii},\ and\
  \citenamefont {del Campo}}]{Pittman17}%
  \BibitemOpen
  \bibfield  {author} {\bibinfo {author} {\bibfnamefont {S.~M.}\ \bibnamefont
  {Pittman}}, \bibinfo {author} {\bibfnamefont {M.}~\bibnamefont {Beau}},
  \bibinfo {author} {\bibfnamefont {M.}~\bibnamefont {Olshanii}}, \ and\
  \bibinfo {author} {\bibfnamefont {A.}~\bibnamefont {del Campo}},\ }\href
  {\doibase 10.1103/PhysRevB.95.205135} {\bibfield  {journal} {\bibinfo
  {journal} {Phys. Rev. B}\ }\textbf {\bibinfo {volume} {95}},\ \bibinfo
  {pages} {205135} (\bibinfo {year} {2017})}\BibitemShut {NoStop}%
\bibitem [{\citenamefont {Tummuru}\ \emph {et~al.}(2017)\citenamefont
  {Tummuru}, \citenamefont {Jain},\ and\ \citenamefont {Khare}}]{Tummuru17}%
  \BibitemOpen
  \bibfield  {author} {\bibinfo {author} {\bibfnamefont {T.~R.}\ \bibnamefont
  {Tummuru}}, \bibinfo {author} {\bibfnamefont {S.~R.}\ \bibnamefont {Jain}}, \
  and\ \bibinfo {author} {\bibfnamefont {A.}~\bibnamefont {Khare}},\ }\href
  {\doibase https://doi.org/10.1016/j.physleta.2017.10.007} {\bibfield
  {journal} {\bibinfo  {journal} {Physics Letters A}\ }\textbf {\bibinfo
  {volume} {381}},\ \bibinfo {pages} {3917 } (\bibinfo {year}
  {2017})}\BibitemShut {NoStop}%
\bibitem [{\citenamefont {Kundu}(1999)}]{Kundu99}%
  \BibitemOpen
  \bibfield  {author} {\bibinfo {author} {\bibfnamefont {A.}~\bibnamefont
  {Kundu}},\ }\href {\doibase 10.1103/PhysRevLett.83.1275} {\bibfield
  {journal} {\bibinfo  {journal} {Phys. Rev. Lett.}\ }\textbf {\bibinfo
  {volume} {83}},\ \bibinfo {pages} {1275} (\bibinfo {year}
  {1999})}\BibitemShut {NoStop}%
\bibitem [{\citenamefont {Calogero}\ and\ \citenamefont
  {Marchioro}(1973)}]{CalogeroMarchioro75}%
  \BibitemOpen
  \bibfield  {author} {\bibinfo {author} {\bibfnamefont {F.}~\bibnamefont
  {Calogero}}\ and\ \bibinfo {author} {\bibfnamefont {C.}~\bibnamefont
  {Marchioro}},\ }\href {\doibase 10.1063/1.1666291} {\bibfield  {journal}
  {\bibinfo  {journal} {Journal of Mathematical Physics}\ }\textbf {\bibinfo
  {volume} {14}},\ \bibinfo {pages} {182} (\bibinfo {year} {1973})}\BibitemShut
  {NoStop}%
\bibitem [{\citenamefont {Kinoshita}\ \emph {et~al.}(2004)\citenamefont
  {Kinoshita}, \citenamefont {Wenger},\ and\ \citenamefont
  {Weiss}}]{Kinoshita04}%
  \BibitemOpen
  \bibfield  {author} {\bibinfo {author} {\bibfnamefont {T.}~\bibnamefont
  {Kinoshita}}, \bibinfo {author} {\bibfnamefont {T.}~\bibnamefont {Wenger}}, \
  and\ \bibinfo {author} {\bibfnamefont {D.~S.}\ \bibnamefont {Weiss}},\ }\href
  {\doibase 10.1126/science.1100700} {\bibfield  {journal} {\bibinfo  {journal}
  {Science}\ }\textbf {\bibinfo {volume} {305}},\ \bibinfo {pages} {1125}
  (\bibinfo {year} {2004})}\BibitemShut {NoStop}%
\bibitem [{\citenamefont {Paredes}\ \emph {et~al.}(2004)\citenamefont
  {Paredes}, \citenamefont {Widera}, \citenamefont {Murg}, \citenamefont
  {Mandel}, \citenamefont {F{\"o}lling}, \citenamefont {Cirac}, \citenamefont
  {Shlyapnikov}, \citenamefont {H{\"a}nsch},\ and\ \citenamefont
  {Bloch}}]{Paredes04}%
  \BibitemOpen
  \bibfield  {author} {\bibinfo {author} {\bibfnamefont {B.}~\bibnamefont
  {Paredes}}, \bibinfo {author} {\bibfnamefont {A.}~\bibnamefont {Widera}},
  \bibinfo {author} {\bibfnamefont {V.}~\bibnamefont {Murg}}, \bibinfo {author}
  {\bibfnamefont {O.}~\bibnamefont {Mandel}}, \bibinfo {author} {\bibfnamefont
  {S.}~\bibnamefont {F{\"o}lling}}, \bibinfo {author} {\bibfnamefont
  {I.}~\bibnamefont {Cirac}}, \bibinfo {author} {\bibfnamefont {G.~V.}\
  \bibnamefont {Shlyapnikov}}, \bibinfo {author} {\bibfnamefont {T.~W.}\
  \bibnamefont {H{\"a}nsch}}, \ and\ \bibinfo {author} {\bibfnamefont
  {I.}~\bibnamefont {Bloch}},\ }\href {\doibase 10.1038/nature02530} {\bibfield
   {journal} {\bibinfo  {journal} {Nature}\ }\textbf {\bibinfo {volume}
  {429}},\ \bibinfo {pages} {277} (\bibinfo {year} {2004})}\BibitemShut
  {NoStop}%
\bibitem [{\citenamefont {Haldane}(1991)}]{Haldane91}%
  \BibitemOpen
  \bibfield  {author} {\bibinfo {author} {\bibfnamefont {F.~D.~M.}\
  \bibnamefont {Haldane}},\ }\href {\doibase 10.1103/PhysRevLett.67.937}
  {\bibfield  {journal} {\bibinfo  {journal} {Phys. Rev. Lett.}\ }\textbf
  {\bibinfo {volume} {67}},\ \bibinfo {pages} {937} (\bibinfo {year}
  {1991})}\BibitemShut {NoStop}%
\bibitem [{\citenamefont {Wu}(1994)}]{Wu94}%
  \BibitemOpen
  \bibfield  {author} {\bibinfo {author} {\bibfnamefont {Y.-S.}\ \bibnamefont
  {Wu}},\ }\href {\doibase 10.1103/PhysRevLett.73.922} {\bibfield  {journal}
  {\bibinfo  {journal} {Phys. Rev. Lett.}\ }\textbf {\bibinfo {volume} {73}},\
  \bibinfo {pages} {922} (\bibinfo {year} {1994})}\BibitemShut {NoStop}%
\bibitem [{\citenamefont {Murthy}\ and\ \citenamefont
  {Shankar}(1994)}]{MurthyShankar94}%
  \BibitemOpen
  \bibfield  {author} {\bibinfo {author} {\bibfnamefont {M.~V.~N.}\
  \bibnamefont {Murthy}}\ and\ \bibinfo {author} {\bibfnamefont
  {R.}~\bibnamefont {Shankar}},\ }\href {\doibase 10.1103/PhysRevLett.73.3331}
  {\bibfield  {journal} {\bibinfo  {journal} {Phys. Rev. Lett.}\ }\textbf
  {\bibinfo {volume} {73}},\ \bibinfo {pages} {3331} (\bibinfo {year}
  {1994})}\BibitemShut {NoStop}%
\bibitem [{\citenamefont {Lieb}\ and\ \citenamefont {Liniger}(1963)}]{LL63}%
  \BibitemOpen
  \bibfield  {author} {\bibinfo {author} {\bibfnamefont {E.~H.}\ \bibnamefont
  {Lieb}}\ and\ \bibinfo {author} {\bibfnamefont {W.}~\bibnamefont {Liniger}},\
  }\href {\doibase 10.1103/PhysRev.130.1605} {\bibfield  {journal} {\bibinfo
  {journal} {Phys. Rev.}\ }\textbf {\bibinfo {volume} {130}},\ \bibinfo {pages}
  {1605} (\bibinfo {year} {1963})}\BibitemShut {NoStop}%
\bibitem [{\citenamefont {Lieb}(1963)}]{L63}%
  \BibitemOpen
  \bibfield  {author} {\bibinfo {author} {\bibfnamefont {E.~H.}\ \bibnamefont
  {Lieb}},\ }\href {\doibase 10.1103/PhysRev.130.1616} {\bibfield  {journal}
  {\bibinfo  {journal} {Phys. Rev.}\ }\textbf {\bibinfo {volume} {130}},\
  \bibinfo {pages} {1616} (\bibinfo {year} {1963})}\BibitemShut {NoStop}%
\bibitem [{\citenamefont {Olshanii}(1998)}]{Olshanii98}%
  \BibitemOpen
  \bibfield  {author} {\bibinfo {author} {\bibfnamefont {M.}~\bibnamefont
  {Olshanii}},\ }\href {\doibase 10.1103/PhysRevLett.81.938} {\bibfield
  {journal} {\bibinfo  {journal} {Phys. Rev. Lett.}\ }\textbf {\bibinfo
  {volume} {81}},\ \bibinfo {pages} {938} (\bibinfo {year} {1998})}\BibitemShut
  {NoStop}%
\bibitem [{\citenamefont {McGuire}(1964)}]{McGuire64}%
  \BibitemOpen
  \bibfield  {author} {\bibinfo {author} {\bibfnamefont {J.~B.}\ \bibnamefont
  {McGuire}},\ }\href {\doibase 10.1063/1.1704156} {\bibfield  {journal}
  {\bibinfo  {journal} {Journal of Mathematical Physics}\ }\textbf {\bibinfo
  {volume} {5}},\ \bibinfo {pages} {622} (\bibinfo {year} {1964})}\BibitemShut
  {NoStop}%
\bibitem [{\citenamefont {Ishikawa}\ and\ \citenamefont
  {Takayama}(1980)}]{Ishikawa80}%
  \BibitemOpen
  \bibfield  {author} {\bibinfo {author} {\bibfnamefont {M.}~\bibnamefont
  {Ishikawa}}\ and\ \bibinfo {author} {\bibfnamefont {H.}~\bibnamefont
  {Takayama}},\ }\href {\doibase 10.1143/JPSJ.49.1242} {\bibfield  {journal}
  {\bibinfo  {journal} {Journal of the Physical Society of Japan}\ }\textbf
  {\bibinfo {volume} {49}},\ \bibinfo {pages} {1242} (\bibinfo {year}
  {1980})}\BibitemShut {NoStop}%
\bibitem [{\citenamefont {Sato}\ \emph {et~al.}(2012)\citenamefont {Sato},
  \citenamefont {Kanamoto}, \citenamefont {Kaminishi},\ and\ \citenamefont
  {Deguchi}}]{Deguchi1}%
  \BibitemOpen
  \bibfield  {author} {\bibinfo {author} {\bibfnamefont {J.}~\bibnamefont
  {Sato}}, \bibinfo {author} {\bibfnamefont {R.}~\bibnamefont {Kanamoto}},
  \bibinfo {author} {\bibfnamefont {E.}~\bibnamefont {Kaminishi}}, \ and\
  \bibinfo {author} {\bibfnamefont {T.}~\bibnamefont {Deguchi}},\ }\href
  {\doibase 10.1103/PhysRevLett.108.110401} {\bibfield  {journal} {\bibinfo
  {journal} {Phys. Rev. Lett.}\ }\textbf {\bibinfo {volume} {108}},\ \bibinfo
  {pages} {110401} (\bibinfo {year} {2012})}\BibitemShut {NoStop}%
\bibitem [{\citenamefont {Sato}\ \emph {et~al.}(2016)\citenamefont {Sato},
  \citenamefont {Kanamoto}, \citenamefont {Kaminishi},\ and\ \citenamefont
  {Deguchi}}]{Deguchi2}%
  \BibitemOpen
  \bibfield  {author} {\bibinfo {author} {\bibfnamefont {J.}~\bibnamefont
  {Sato}}, \bibinfo {author} {\bibfnamefont {R.}~\bibnamefont {Kanamoto}},
  \bibinfo {author} {\bibfnamefont {E.}~\bibnamefont {Kaminishi}}, \ and\
  \bibinfo {author} {\bibfnamefont {T.}~\bibnamefont {Deguchi}},\ }\href
  {\doibase 10.1088/1367-2630/18/7/075008} {\bibfield  {journal} {\bibinfo
  {journal} {New Journal of Physics}\ }\textbf {\bibinfo {volume} {18}},\
  \bibinfo {pages} {075008} (\bibinfo {year} {2016})}\BibitemShut {NoStop}%
\bibitem [{\citenamefont {Bracewell}(2000)}]{Bracewell00}%
  \BibitemOpen
  \bibfield  {author} {\bibinfo {author} {\bibfnamefont {R.~N.}\ \bibnamefont
  {Bracewell}},\ }\href@noop {} {\emph {\bibinfo {title} {The Fourier Transform
  and its Applications}}},\ \bibinfo {edition} {3rd}\ ed.\ (\bibinfo
  {publisher} {McGraw-Hill},\ \bibinfo {year} {2000})\BibitemShut {NoStop}%
\bibitem [{\citenamefont {Beau}\ \emph {et~al.}(2020)\citenamefont {Beau},
  \citenamefont {Pittman}, \citenamefont {Astrakarchik},\ and\ \citenamefont
  {del Campo}}]{BeauPittman20}%
  \BibitemOpen
  \bibfield  {author} {\bibinfo {author} {\bibfnamefont {M.}~\bibnamefont
  {Beau}}, \bibinfo {author} {\bibfnamefont {S.~M.}\ \bibnamefont {Pittman}},
  \bibinfo {author} {\bibfnamefont {G.}~\bibnamefont {Astrakarchik}}, \ and\
  \bibinfo {author} {\bibfnamefont {A.}~\bibnamefont {del Campo}},\ }\href@noop
  {} {\bibfield  {journal} {\bibinfo  {journal} {TBS}\ } (\bibinfo {year}
  {2020})}\BibitemShut {NoStop}%
\bibitem [{\citenamefont {Polychronakos}(1992)}]{Polychronakos92}%
  \BibitemOpen
  \bibfield  {author} {\bibinfo {author} {\bibfnamefont {A.~P.}\ \bibnamefont
  {Polychronakos}},\ }\href {\doibase 10.1103/PhysRevLett.69.703} {\bibfield
  {journal} {\bibinfo  {journal} {Phys. Rev. Lett.}\ }\textbf {\bibinfo
  {volume} {69}},\ \bibinfo {pages} {703} (\bibinfo {year} {1992})}\BibitemShut
  {NoStop}%
\bibitem [{\citenamefont {Muriel}(1976)}]{Muriel76}%
  \BibitemOpen
  \bibfield  {author} {\bibinfo {author} {\bibfnamefont {A.}~\bibnamefont
  {Muriel}},\ }\href {\doibase https://doi.org/10.1016/0375-9601(76)90365-0}
  {\bibfield  {journal} {\bibinfo  {journal} {Physics Letters A}\ }\textbf
  {\bibinfo {volume} {56}},\ \bibinfo {pages} {343 } (\bibinfo {year}
  {1976})}\BibitemShut {NoStop}%
\bibitem [{\citenamefont {Toda}(1967)}]{Toda67}%
  \BibitemOpen
  \bibfield  {author} {\bibinfo {author} {\bibfnamefont {M.}~\bibnamefont
  {Toda}},\ }\href {\doibase 10.1143/JPSJ.22.431} {\bibfield  {journal}
  {\bibinfo  {journal} {Journal of the Physical Society of Japan}\ }\textbf
  {\bibinfo {volume} {22}},\ \bibinfo {pages} {431} (\bibinfo {year}
  {1967})}\BibitemShut {NoStop}%
\bibitem [{\citenamefont {Cazorla}\ and\ \citenamefont
  {Boronat}(2007)}]{Cazorla07}%
  \BibitemOpen
  \bibfield  {author} {\bibinfo {author} {\bibfnamefont {C.}~\bibnamefont
  {Cazorla}}\ and\ \bibinfo {author} {\bibfnamefont {J.}~\bibnamefont
  {Boronat}},\ }\href {\doibase 10.1088/0953-8984/20/01/015223} {\bibfield
  {journal} {\bibinfo  {journal} {Journal of Physics: Condensed Matter}\
  }\textbf {\bibinfo {volume} {20}},\ \bibinfo {pages} {015223} (\bibinfo
  {year} {2007})}\BibitemShut {NoStop}%
\bibitem [{\citenamefont {Cazorla}\ \emph {et~al.}(2009)\citenamefont
  {Cazorla}, \citenamefont {Astrakharchik}, \citenamefont {Casulleras},\ and\
  \citenamefont {Boronat}}]{Cazorla09}%
  \BibitemOpen
  \bibfield  {author} {\bibinfo {author} {\bibfnamefont {C.}~\bibnamefont
  {Cazorla}}, \bibinfo {author} {\bibfnamefont {G.~E.}\ \bibnamefont
  {Astrakharchik}}, \bibinfo {author} {\bibfnamefont {J.}~\bibnamefont
  {Casulleras}}, \ and\ \bibinfo {author} {\bibfnamefont {J.}~\bibnamefont
  {Boronat}},\ }\href {\doibase 10.1088/1367-2630/11/1/013047} {\bibfield
  {journal} {\bibinfo  {journal} {New Journal of Physics}\ }\textbf {\bibinfo
  {volume} {11}},\ \bibinfo {pages} {013047} (\bibinfo {year}
  {2009})}\BibitemShut {NoStop}%
\bibitem [{\citenamefont {Khare}\ and\ \citenamefont {Ray}(1997)}]{KhareRay97}%
  \BibitemOpen
  \bibfield  {author} {\bibinfo {author} {\bibfnamefont {A.}~\bibnamefont
  {Khare}}\ and\ \bibinfo {author} {\bibfnamefont {K.}~\bibnamefont {Ray}},\
  }\href {\doibase https://doi.org/10.1016/S0375-9601(97)00233-8} {\bibfield
  {journal} {\bibinfo  {journal} {Physics Letters A}\ }\textbf {\bibinfo
  {volume} {230}},\ \bibinfo {pages} {139 } (\bibinfo {year}
  {1997})}\BibitemShut {NoStop}%
\bibitem [{\citenamefont {Girardeau}\ \emph {et~al.}(2004)\citenamefont
  {Girardeau}, \citenamefont {Nguyen},\ and\ \citenamefont
  {Olshanii}}]{Girardeau04}%
  \BibitemOpen
  \bibfield  {author} {\bibinfo {author} {\bibfnamefont {M.}~\bibnamefont
  {Girardeau}}, \bibinfo {author} {\bibfnamefont {H.}~\bibnamefont {Nguyen}}, \
  and\ \bibinfo {author} {\bibfnamefont {M.}~\bibnamefont {Olshanii}},\ }\href
  {\doibase https://doi.org/10.1016/j.optcom.2004.09.079} {\bibfield  {journal}
  {\bibinfo  {journal} {Optics Communications}\ }\textbf {\bibinfo {volume}
  {243}},\ \bibinfo {pages} {3 } (\bibinfo {year} {2004})}\BibitemShut
  {NoStop}%
\end{thebibliography}%

\end{document}